\documentclass[mathleft
]{an}
\usepackage{graphicx}
\usepackage{times}
\begin{document}

\Pagespan{789}{}
\Yearpublication{2006}%
\Yearsubmission{2005}%
\Month{11}%
\Volume{999}%
\Issue{88}%

\title{Characteristics of 100+ Kepler Asteroseismic Targets from Ground-Based 
Observations\,\thanks{The data used in this paper have been obtained at the
Wroc\l{}aw University Observatory in Bia\l{}k\'ow, the Osservatorio Astrofisico di 
Catania, the F.\ L.\ Whipple Observatory, Mount Hopkins, Arizona, the Oak Ridge Observatory, 
Harvard, Massachusetts, and the MMT.}}

\author{J.\ Molenda-\.Zakowicz\inst{1}\fnmsep\thanks{\email{molenda@astro.uni.wroc.pl}\newline}, 
M.\ Jerzykiewicz\inst{1}, A.\ Frasca\inst{2}, G.\ Catanzaro\inst{2}, G.\ Kopacki\inst{1}, 
and D.W.\ Latham\inst{3}
}
\titlerunning{Characteristics of 100+ Kepler Asteroseismic Targets from Ground-Based
Observations}
\authorrunning{J.\ Molenda-\.Zakowicz et al.}
\institute{
Instytut Astronomiczny Uniwersytetu Wroc\l{}awskiego, ul.\ Kopernika 11, 51-622 Wroc\l{}aw, Poland
\and 
INAF--Osservatorio Astrofisico di Catania, Via Sofia 78, 95123 Catania, Italy
\and 
Harvard Smithsonian Center for Astrophysics, Cambridge, Massachusetts, USA
}

\received{1 April 2010}
\accepted{---}
\publonline{later}


\abstract{%
We present results of our 5-years-long program of ground-based spectroscopic and photometric observations 
of individual Kepler asteroseismic targets and the open clusters NGC\,6866 and NGC\,6811 from the Kepler 
field of view. We determined the effective temperature, surface gravity, metallicity, the projected 
rotational velocity and the radial velocity of 119 Kepler asteroseismic targets for which we acquired high-resolution 
spectra. For many of these stars the derived atmospheric parameters agree with $T_{\rm eff}$, $\log g$, and 
[Fe/H] from the Kepler Input Catalog (KIC) to within their error bars. Only for stars hotter than 7000\,K we 
notice significant differences between the effective temperature derived from spectroscopy and $T_{\rm eff}$ given 
in the KIC. For 19 stars which we observed photoelectrically, we measured the interstellar reddening and we found 
it to be negligible. Finally, our discovery of the $\delta$ Sct and $\gamma$ Dor pulsating stars in the open 
cluster NGC\,6866 allowed us to discuss the frequency of the occurrence of $\gamma$ Dor stars in the 
open clusters of different age and metallicity and show that there are no correlations between these parameters.
}
\maketitle

\section{Introduction}

Our program of ground-based spectroscopic and photometric observations of stars selected for the asteroseismic
targets for the Kepler space telescope by the Kepler Asteroseismic Science Consortium KASC\footnote{Kepler 
Asteroseismic Science Consortium (KASC) is a group of collaborating scientists and/or institutions established 
to accomplish the activities of the Kepler Asteroseismic Investigation (KAI), represented by Ronald Gilliland
(see http://astro.phys.au.dk/KASC).} was started in 2005 at the Osservatorio
Astrofisico di Catania, OACt, (the {\it M.G. Fracastoro\/} station, Mt.\ Etna, Italy) and is continued since then.
Apart from the OACt, we perform spectroscopic and photometric observations of Kepler asteroseismic targets
at the F.\ L.\ Whipple Observatory, FLWO, (Mount Hopkins, Arizona, USA), and the Astrophysical Observatory of 
the University of Wroc\l{}aw in Bia\l{}k\'ow (Poland). 

At the OACt, we use a 91-cm telescope, at FLWO, a 1.5-m telescope, and at the Bia\l{}k\'ow Observatory, a 60-cm 
telescope. We make use also of the archival data collected at the 1.5-m telescope at the Oak Ridge 
Observatory, ORO, (Harvard, Massachusetts, USA) and at the Multiple Mirror Telescope, MMT, before it was 
converted to the monolithic 6.5-m mirror. 

Using the spectroscopic data acquired at OACt, FLWO, ORO, and MMT, we aim at the determination of the atmospheric 
parameters of the program stars, i.e., the effective temperature, $T_{\rm eff}$, surface gravity, $\log g$, 
and metallicity, $\rm [Fe/H]$, and measuring the projected rotational velocity, $v\sin i$, and the radial velocity, 
$v_r$, of the stars. 

At the Bia\l{}k\'ow Observatory, we perform photometric time-series observations of the open clusters 
NGC\,6866 and NGC\,6811 in  the Kepler field of view, aiming at the discovery of new pulsating stars, and the 
determination of the degree of the modes of their pulsations.

Both sites, the Bia\l{}k\'ow Observatory and the OACt, took part in the international multi-site photometric
campaign on NGC\,6866 launched in 2009, and will take part in a similar campaign on NGC\,6811 which will be launched in 2010
(for the details, see Uytterhoeven et al. 2010b.)

\begin{figure}
\includegraphics[width=60mm,height=80mm,angle=270]{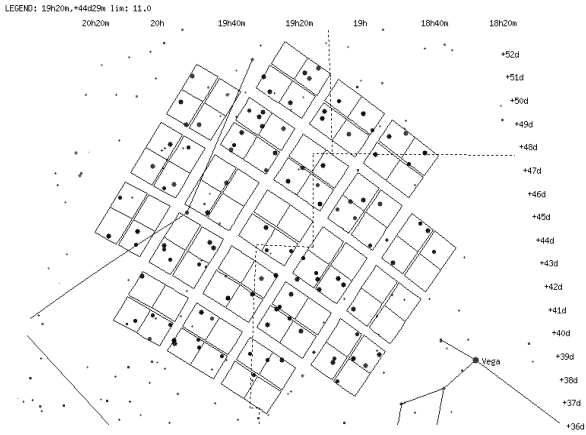}
\caption{The Kepler asteroseismic targets observed at the Osservatorio Astrofisico di Catania, 
the F.\ L.\ Whipple Observatory, and the Oak Ridge Observatory. Indicated are the borders of the 42 Kepler CCDs
and the borders of the constellations.}
\label{targets}
\end{figure}

\section{Radial velocity and the projected rotational velocity}

When analyzing the spectrograms with the aim of deriving the atmospheric parameters, we measured 
the projected rotational velocity of the program stars and their radial velocity. As expected, 
we found the F, G, and K type stars to be slow rotators having $v\sin i$ typically below 5 km\,s$^{-1}$.
The early-type stars discussed by Catanzaro et al.\ (2010) rotate significantly faster and although several
of them have $v\sin i < 10$ km\,s$^{-1}$, this low projected rotational velocity may be due to the 
orientation of the axis of the rotation of the star in the sky.

From the analysis of the radial velocities of the program stars, we discovered six single-lined, SB1, and four 
double-lined, SB2, spectroscopic binaries (see Molenda-\.Zakowicz et al.\ 2007, 2010; Molenda-\.Zakowicz, Frasca 
\& Latham 2008; Catanzaro et al.\ 2010; Frasca et al. 2010). For two of the new SB1 systems and one new SB2 star 
we calculated the systems' mass functions and the orbital solutions, respectively. More spectroscopic and photometric 
observations are needed to derive the mass functions and orbital solutions for the remaining stars, and to check 
which of these systems are eclipsing.

In Table \ref{spectro}, we give the number of the spectrograms acquired for each program star at the OACt, 
FLWO, ORO and MMT.

\begin{table*}
\centering
\caption{The number of the spectrograms acquired for each program star at the Osservatorio Astrofisico di Catania, OACt,
the F.\ L.\ Whipple Observatory, FLWO, the Oak Ridge Observatory, ORO, and the Multiple Mirror Telescope, MMT.}
\label{spectro}
{\tiny
\begin{tabular}{rrrrr|rrrrr|rrrrr}\hline
\noalign{\smallskip}
KIC    & OACt & FLWO & ORO & MMT & KIC    & OACt & FLWO & ORO & MMT & KIC    & OACt & FLWO & ORO & MMT\\ 
\noalign{\smallskip}
\hline
\noalign{\smallskip}
 2696947 &    1 &   1  & --  & -- &  	 6945099 &    1 &  --  & --  & -- &  	10068307 &    1 &  --  & --  & --\\
 2991548 &    1 &  --  & --  & -- &  	 6976475 &    1 &   1  & --  & -- &  	10124866 &    1 &  49  & --  &  6\\
 3347643 &    2 &  --  & --  & -- &  	 7022603 &    1 &  --  & --  & -- &  	10131030 &    1 &  --  & --  & --\\
 3425374 &    1 &   1  & --  & -- &  	 7341231 &    1 &  --  & --  & -- &  	10162436 &    1 &  --  & --  & --\\
 3632418 &    1 &  --  & --  & -- &  	 7374855 &    1 &  --  & --  & -- &  	10187831 &    1 &  --  & --  & --\\
 3641446 &    1 &  --  & --  & -- &  	 7548061 &    1 &  --  & --  & -- &  	10454113 &    1 &  --  & --  & --\\
 3644223 &    2 &  --  & --  & -- &  	 7599132 &    1 &  --  & --  & -- &  	10513837 &    3 &   1  & --  & --\\ 
 3730953 &    1 &  --  & --  & -- &  	 7730305 &    2 &  --  & --  & -- &  	10532461 &    2 &  --  &  1  & --\\
 3733735 &    1 &  --  &  1  & -- &  	 7747078 &    2 &  --  & --  & -- &  	10604429 &    3 &  --  & --  & --\\
 3747220 &    1 &  --  & --  & -- &  	 7820638 &    1 &  --  & --  & -- &  	10677958 &    1 &  --  & --  & --\\
 3830233 &    1 &  --  & --  & -- &  	 7841024 &    1 &  --  & --  & -- &  	10735274 &    1 &  --  & --  & --\\
 3858884 &    1 &  --  & --  & -- &  	 7898839 &    1 &  --  &  1  & -- &  	10748390 &    2 &   4  &   1 & --\\
 3956527 &    1 &   4  & --  & -- &  	 7944142 &    2 &  --  & --  & -- &  	10960750 &    1 &  --  & --  & --\\
 4150611 &    5 &  --  & --  & -- &  	 7978223 &    3 &  --  & --  & -- &  	11013201 &    1 &  --  & --  & --\\ 
 4276892 &    1 &  --  & --  & -- &  	 7985370 &    1 &  --  & --  & -- &  	11018874 &    1 &  --  &  1  & --\\
 4484238 &    1 &  --  &  1  & -- &  	 8037268 &    2 &   2  &   1 & -- &  	11031993 &    2 &  33  & 36  &  3\\
 4574610 &    2 &  --  & --  & -- &  	 8264549 &    1 &   1  & --  & -- &  	11070918 &    2 &   2  & --  & --\\
 4581434 &    1 &  --  & --  & -- &  	 8343931 &    2 &  --  & --  & -- &  	11134456 &    1 &  --  & --  & --\\
 4681323 &    1 &  --  & --  & -- &  	 8379927 &    1 &   3  & 40  & -- &  	11189959 &    1 &  --  & --  & --\\
 4818496 &    1 &  --  & --  & -- &  	 8389948 &    1 &  --  & --  & -- &  	11253226 &    2 &  --  & --  & --\\
 4914923 &   11 &   5  & --  & -- &  	 8429280 &    2 &  --  & --  & -- &  	11255615 &    2 &  --  & --  & --\\
 5184732 &    1 &  --  & --  & -- &  	 8539201 &    2 &  --  & --  & -- &  	11342410 &    2 &  --  & --  & --\\
 5206997 &    1 &  --  & --  & -- &  	 8547390 &    1 &  --  & --  & -- &  	11402951 &    1 &  --  & --  & --\\
 5304891 &    1 &  --  & --  & -- &  	 8561664 &    1 &  --  &  1  & -- &  	11495120 &    1 &  --  & --  & --\\
 5371516 &    1 &  --  & --  & -- &  	 8677933 &    1 &  --  & --  & -- &  	11498538 &    1 &  --  & --  & --\\
 5442047 &    1 &  --  & --  & -- &  	 8740371 &    1 &  --  & --  & -- &  	11506859 &    2 &  --  &  1  & --\\
 5557932 &    2 &  --  &  1  & -- &  	 8894567 &    2 &   2  &   1 & -- &  	11551430 &    1 &  --  & --  & --\\
 5631061 &    1 &  --  & --  & -- &  	 8940939 &    1 &  --  & --  & -- &  	11560431 &    7 &  --  & --  & --\\
 5701829 &    1 &  --  & --  & -- &  	 9139151 &    2 &  --  & --  & -- &  	11708170 &    1 &  --  &  1  & --\\
 5774694 &    1 &   1  & 23  & -- &  	 9139163 &    2 &  --  & --  & -- &  	11709006 &    2 &  --  &  1  & --\\
 5786771 &    1 &  --  & --  & -- &  	 9145955 &    1 &  --  & --  & -- &  	11754082 &    3 &  --  & --  & --\\
 6128830 &    1 &  --  & --  & -- &  	 9204877 &    1 &   2  & --  & -- &  	11762256 &    1 &  --  & --  & --\\
 6278762 &    2 &   6  & 14  &  1 &  	 9206432 &    1 &  --  & --  & -- &  	11775000 &    2 &  --  & --  & --\\
 6285677 &    1 &  --  & --  & -- &  	 9307354 &    2 &   2  & --  & -- &  	12250891 &    1 &  --  & --  & --\\
 6432054 &    1 &  --  & --  & -- &  	 9605196 &    2 &  --  & --  & -- &  	12253106 &    1 &  --  & --  & --\\
 6590668 &    1 &  --  & --  & -- &   	 9641031 &    2 &  12  & --  & -- &  	12258514 &    2 &  --  &  1  & --\\
 6766118 &    1 &  --  & --  & -- &  	 9663677 &    1 &  --  & --  & -- &  	12317678 &    1 &  --  & --  & --\\
 6769635 &    1 &  --  & --  & -- &  	 9705687 &    1 &  --  & --  & -- &  	12352180 &    2 &   4  & --  & --\\
 6848529 &    1 &  --  & --  & -- &  	 9715099 &    1 &  --  & --  & -- &  	12453925 &    1 &  --  & --  & --\\
 6862114 &    1 &  --  & --  & -- &  	10010623 &    2 &  --  & --  & -- &  \\	
\hline
\end{tabular}
}
\end{table*}

\section{Atmospheric parameters}

We measured the effective temperature, $T_{\rm eff}$, the surface gravity, $\log g$, and the 
metallicity, $\rm [Fe/H]$, of 119 stars selected for asteroseismic targets for Kepler. There is one
sub-dwarf in this sample, HIP\,92775; all the remaining stars have solar metallicity or are slightly 
metal-deficient (see Molenda-\.Zakowicz et al.\ 2007, 2010; Molenda-\.Zakowicz, Frasca \& Latham 2008; 
Catanzaro et al.\ 2010; Frasca et al. 2010.)

\begin{figure}
\includegraphics[width=200mm,height=80mm,angle=-90]{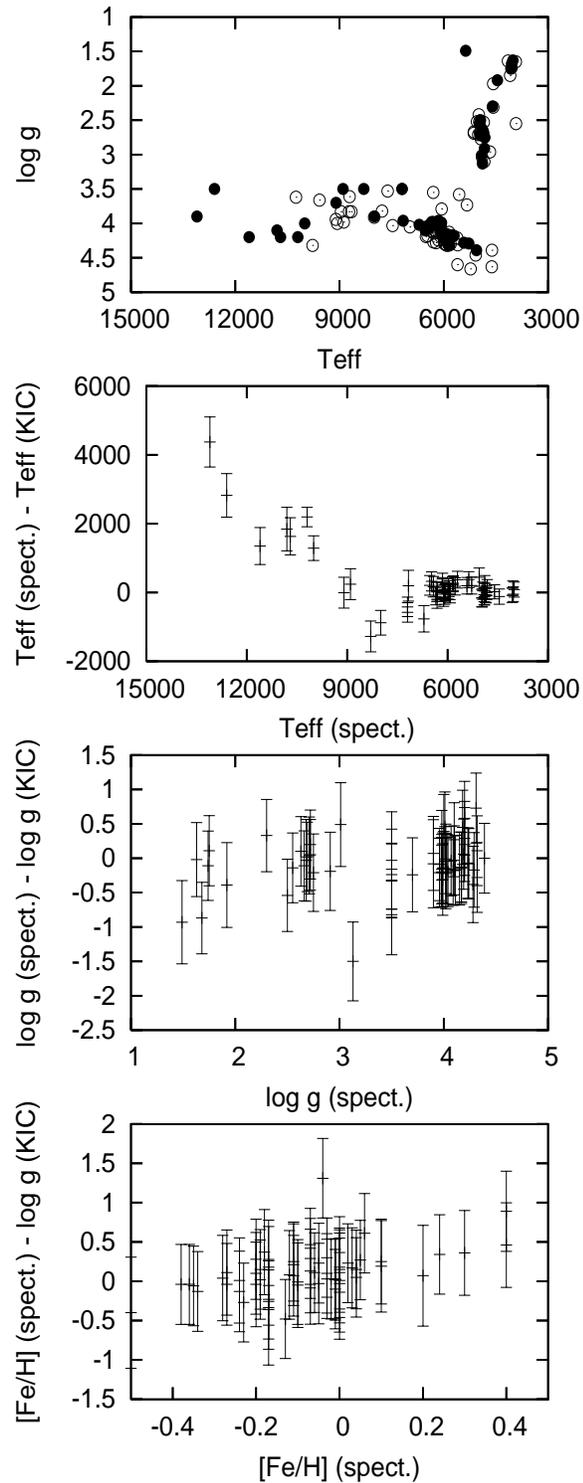}
\caption{{\it Top:} The $T_{\rm eff} - \log g$ diagram for the program stars plotted with the use of $T_{\rm eff}$
and $\log g$ from the Kepler Input Catalog KIC (circles) and the values derived from spectroscopy (dots). 
{\it The next panels from top to bottom:} The differences between $T_{\rm eff}$, $\log g$ and [Fe/H] 
derived from spectroscopy and those given in the KIC.}
\label{atm_par}
\end{figure}

In the top panel of Fig.\ \ref{atm_par}, we plot the program stars in the $T_{\rm eff}$ -- $\log g$ diagram 
constructed with the use of the parameters derived from spectroscopy (dots) and those from the Kepler 
Input Catalog\footnote{http://archive.stsci.edu/} KIC (circles) derived from the photometric observations 
acquired in the Sloan filters. The next panels show the differences between the $T_{\rm eff}$, 
$\log g$ and [Fe/H] derived from spectroscopy and those given in the KIC.

As can be seen in the figure, the metallicity derived from spectroscopy and given in the KIC agree satisfactory
to wi\-thin their error bars. We note, however, that the latter are quite large as the uncertainty of [Fe/H] in 
the KIC is equal to $\pm0.5$ dex. 

The agreement of the values of the surface gravity derived form spectroscopy and given in the KIC is 
satisfactory for main-sequence stars. For more evolved stars the differences between $\log g$ can be as high 
as 1.5 dex. Again, the uncertainty of the determination of $\log g$ in the KIC is equal to $\pm0.5$ dex. 

The largest discrepancies occur for 
the effective temperatures of stars hotter than 7000\,K. The origin of these discrepancies is not clear. The
uncertainty of $T_{\rm eff}$ in the KIC is $\pm200$ K in the whole discussed range of $T_{\rm eff}$.

\section{Interstellar reddening}
In Molenda-\.Zakowicz, Jerzykiewicz \& Frasca (2009a), we report deriving the interstellar reddening for 29 stars 
in the Kepler field of view; 19 of these stars are asteroseismic targets for Kepler. We find that, contrary 
to the information in the KIC, these stars are not reddened while the KIC gives $E(B-V)$ ranging from 0.01 to 0.06 
mag for nine of our program stars. We will continue measuring the reddening of the Kepler asteroseismic targets
using the data which will be collected for stars selected for photometric observations at several observing sites 
in 2010. For the details of our observing programme, we refer to Uytterhoeven et al.\ (2010a).

\section{Open Clusters}

In 2007, 2008 and 2009 at the OACt and the Bia\l{}k\'ow Observatory, we observed two open clusters in the  Kepler field of 
view: NGC\,6811 and NGC\,6866. 

The CCD multicolour observations of NGC\,6866 acquired at the Bia\l{}k\'ow Observatory in 2007 allowed us 
to discover 19 variable stars of different types, including three pulsating stars of the $\delta$ Sct type, and 
two, of the $\gamma$ Dor type (Molenda-\.Zakowicz et al.\ 2009b). All these five stars have been included in 
the list of Kepler asteroseismic targets. 

The discovery of $\gamma$ Dor stars in NGC\,6866 allowed us to discuss the properties of open clusters of 
different age and metallicity which host $\gamma$ Dor stars (see Molenda-\.Zakowicz et al.\ 2009). We showed 
that there is no relation between these two parameters and the number of $\gamma$ Dor stars in the clusters.

In 2009, the open cluster NGC\,6866 was a subject of a multi-site photometric campaign; in 2010, a similar campaign will 
be launched for NGC 6811 (see Uytterhoeven et al. 2010b). The aim of these campaigns is to determine the differences in the 
amplitudes and phases of the pulsating stars measured in the $B$ and $V$ Johnson filters, and to derive the degree of the excited 
modes of pulsations, $l$, which will help computing the asteroseismic models of these stars.

\acknowledgements
This work was supported by MNiSW grant N203 014 31/2650 and the University of 
Wroc{\l}aw grants No 2646 /W/IA/06 and 2793/W/IA/07.

\end{document}